\documentclass[reprint,
aps,
pra,
]{revtex4-2}
\usepackage{graphicx}
\usepackage{dcolumn}
\usepackage{bm}
\usepackage{booktabs} 
\usepackage{adjustbox} 

\begin{document}

\title{Emission from driven atoms in collective strong coupling with an optical cavity}
 \author{V.R. Thakar}%
 \email{vardhanr@rri.res.in}

 \author{Arun Bahuleyan}%

 \author{V. I. Gokul}%

 \author{S. P. Dinesh}%

 \author{S. A. Rangwala}%
  \email{sarangwala@rri.res.in}

 \affiliation{%
  Raman Research Institute, C. V. Raman Avenue, Sadashivanagar, Bangalore 560080, India \\}

\begin{abstract}

We study self sustained cavity emission from driven atoms in collective strong coupling.
The cavity emission occurs over a wide range of atom-cavity and drive laser detunings without any external input to the cavity mode.
Second order correlation measurements ($g^2(\tau)$), further reveal unanticipated phenomenon in the observed cavity emission such as, (a) damped oscillations at two frequencies and (b) significantly distinct $g^2(\tau)$ for different polarization components.
The intricate relation between cavity emission intensity, drive laser detuning and atom-cavity detunings is explained.
A possible mechanism for the damped oscillations with two frequency components in $g^2(\tau)$ is suggested.
Measurements show the existence of two separate polarization decoupled mechanisms with distinct photon statistics, through which energy is transferred from the drive field to the cavity field. The statistical properties and mechanisms underlying cavity emission, as presented in this work, are expected to provide valuable insights for extending non-destructive detection techniques to the regime of collective strong coupling.

\end{abstract}

\maketitle

\section{Introduction}

Atoms in free space interacting with a near-resonant drive laser radiate via coherent and incoherent scattering. The experimentally observed spectrum of this scattered light exhibits a characteristic three peak structure called the Mollow triplet.
The central peak, appearing at the drive laser frequency ($\omega_l$), predominantly comprises of coherently scattered photons.
Symmetric sideband peaks are observed at frequencies $\omega_l \pm \Omega_R$, where $\Omega_R$ is the generalized Rabi frequency, primarily comprising of incoherently scattered photons \cite{Mollow1, Mollow3, Mollow2, Grynberg}.
When an ensemble of driven atoms is confined within the spatial mode of an optical cavity, both the spatial and spectral properties of the scattered light are affected \cite{purcell, Kleppner_Purcell, Mossberg}.

For cavity-coupled atoms, interactions are characterized by the atom-cavity coupling strength ($g_0$), the atomic excited state decay rate ($\Gamma$) and the cavity photon loss rate ($\kappa$) \cite{Jaynes_Cummings_SC, Rempe_VRS, BEC_cav, Kasevich, Brecha_Carmichael_1, carmichael, Zhou_Swain}.
Typically, in cavities with large mode volume and low finesse, $g_0 \ll \Gamma, \kappa$ and hence the system is considered to be in weak coupling regime \cite{carmichael, Zhou_Swain, Mossberg}.
In this regime, Lezama et. al. have shown in atomic beam experiments where, as the resonant cavity frequency ($\omega_c$) is tuned close to the central Mollow peak at the drive laser frequency ($\omega_c \approx \omega_l$), emission can be observed along the cavity axis without any external input to the cavity mode \cite{Mossberg}.
Additionally, at sufficiently high optical densities, cavity emission was also observed when the cavity was tuned close to the red sideband of the Mollow triplet ($\omega_c \approx \omega_l-\Omega_R$).

\begin{figure}[t]
    \begin{centering}
    \includegraphics[width=0.98\linewidth, keepaspectratio]{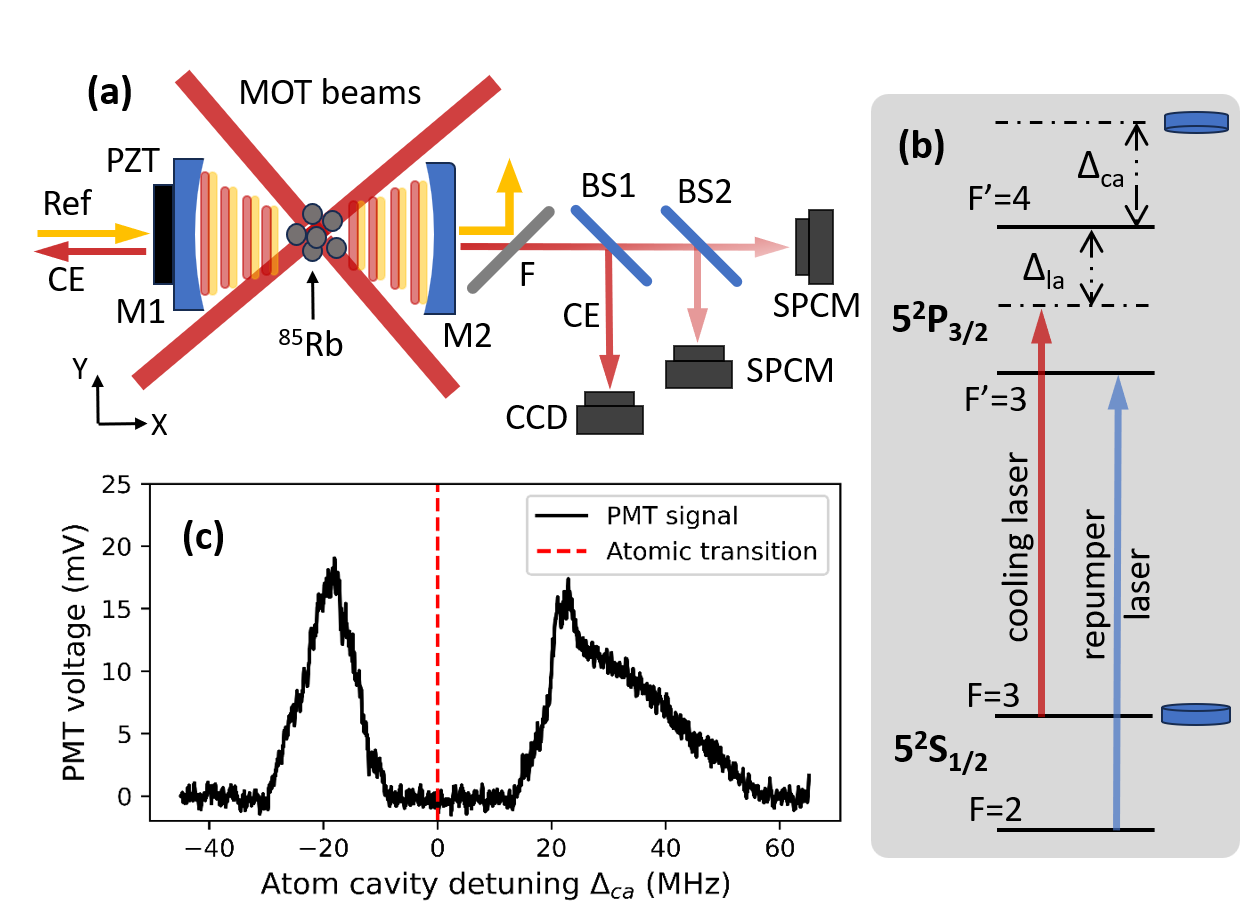}
        \label{Red_Blue_Peaks}
        \caption{(a) Schematic of the experimental setup: mirrors M1, M2 with the piezo electric transducer (PZT) form the tunable cavity. A 767 nm laser (Ref) stabilizes the cavity. MOT beams trap the atoms and act as the drive. An optical band-pass filter (F) is used to block the stabilization laser while allowing the cavity emission (CE) to pass through. Beam splitter (BS1) is used to monitor the spatial profile of the emission on a camera (CCD). The transmitted light is coupled to a multi-mode fiber and fed to a fiber based beam splitter (BS2) whose outputs are recorded on single photon counting modules (SPCM) for $g^2(\tau)$ measurement. (b) Relevant energy levels of $^{85}$Rb and detunings for the experiment are shown. (c) Profiles of observed emission as the cavity is scanned across the $F=3 \leftrightarrow F'=4$ transition. Here, $N_c \approx 20000$, $\Delta_{la} \approx -2\Gamma$ and drive power is 25 mW.}
	\end{centering}
\end{figure}

A distinct regime of atom cavity interactions is achieved when a large number of atoms collectively interact with the cavity, such that $g_0 \sqrt{N_c} > \Gamma, \kappa$, where $N_c$ is the effective number of atoms coupled to the cavity \cite{Tavis, Agarwal_vrs}.
This collective strong coupling alters the resonant frequency of the cavity mode producing vacuum Rabi splitting (VRS) \cite{ Brecha_Carmichael_1, Tavis, Agarwal_vrs, Zhu_Line, SAR_Tridib_Nc, SAR4, SAR_Goku}.
Hence in this regime, unlike Lezama et. al., no emission is observed at $\omega_c \approx \omega_l$, or $\omega_c \approx \omega_l-\Omega_R$.
Instead, here, the cavity emission occurs in two cases: (a) when the cavity is red shifted relative to the atomic transition ($\omega_c < \omega_a$; $\omega_a$ is the angular frequency of the atomic transition) so that the red side VRS peak matches the gain in the red sideband of the Mollow spectrum; and (b) when the cavity is blue detuned ($\omega_c > \omega_a $) to ensure the red side VRS peak matches the central peak of the Mollow spectrum \cite{SAR_Rahul}.
The schematic of the experimental implementation and corresponding cavity emissions are shown in Fig. 1.

Case (a) has been studied recently in \cite{SAR_Arun}. In this work we study case (b).
We investigate the range of drive laser detuning ($\Delta_{la} = (\omega_l - \omega_a) / 2 \pi$) and the atom-cavity detuning ($\Delta_{ca} = (\omega_c - \omega_a) / 2 \pi$) over which the blue detuned cavity emission is observed.
At a fixed drive detuning, $g^2(\tau)$ is measured at various $\Delta_{ca}$ to understand the statistical properties of the cavity emission.
Unexpected behavior such as, damped oscillations at two distinct frequencies, which vary with $\Delta_{ca}$ and distinct photon statistics for different polarization components of the cavity emission are seen.
The observed dependence of cavity emission on $\Delta_{la}$ and $\Delta_{ca}$ is discussed with a simple model of a driven two level system interacting with a cavity.
A possible mechanism for the existence of two frequencies observed in the $g^2(\tau)$ is suggested.

\section{Experimental Procedure and Results}

Our experimental setup consists of a dilute cloud of $^{85}$Rb atoms confined in a magneto-optical trap (MOT) co-centered with a moderate finesse optical Fabry-Perot cavity as shown in Fig. 1(a). The relevant energy levels are shown in Fig. 1(b). Further details of the experimental setup can be found in previous work \cite{SAR_Tridib_Nc, SAR_Rahul}.
The cavity emission from the output cavity mirror is passed through a 50:50 beam splitter. One arm is used to monitor the spatial profile of cavity emission while the other is collected in multi-mode fiber and fed to a second order correlation measurement setup \cite{HBT, Shafi}.
The correlation setup consists of a multi-mode fiber based beam splitter and two single photon detectors.
Over $10^7$ timestamps and detector IDs are recorded and the $g^2(\tau)$ is constructed using the multi-stop multi-start algorithm available in correlation functions library of Python programming language \cite{Python}.
\begin{figure}[t]
    \begin{center}
    \includegraphics[width=0.94\linewidth, keepaspectratio]{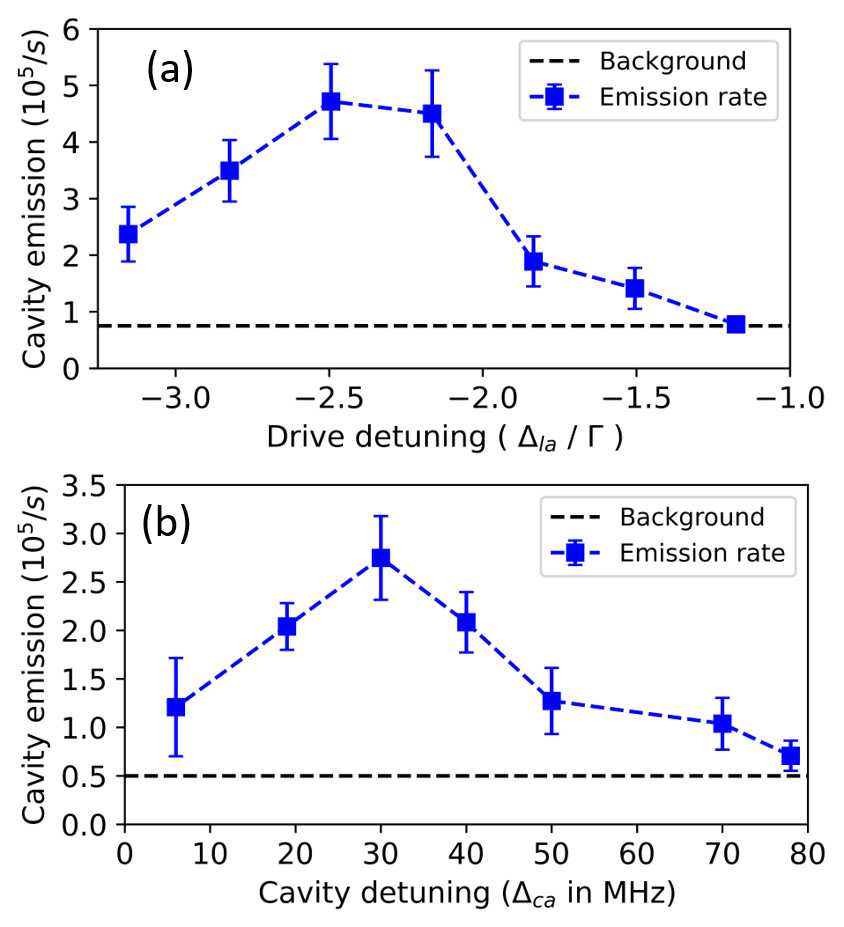}
        \label{Effect_of_drive,cavity}
        \caption{(a) Measured cavity emission intensity achieved by maximization for each $\Delta_{la}$. The emission intensity is recorded over a duration of 100 seconds. The average rate and standard deviation are shown as blue squares with error bars. The total background counts including the dark counts are close to $80\times 10^3$ per second, shown as dashed black line. The drive laser power is kept constant at 25 mW. (b) At 25 mW drive laser power, $\Delta_{la} \approx -3\Gamma$ with $N_c \approx 23000$, emission is observed over nearly 80 MHz range of $\Delta_{ca}$. At each cavity position, the emission intensity is recorded for over 100 seconds. The average and standard deviation are presented as blue squares with error bars. The total background counts are $50\times 10^3$ per second.}
    \end{center}
\end{figure}

In the absence of an external input to the cavity mode, the observed cavity emission must originate from the photons scattered by the MOT atoms. The total scattering rate and the fraction of coherent and incoherent scattering from a driven two level system can be controlled by varying the drive laser frequency. Hence, we study the dependence of the cavity emission on the drive laser detuning for $-3.25 \Gamma \leq \Delta_{la} \leq -1.25 \Gamma$, varied in steps of $0.33 \Gamma$.
At each value of $\Delta_{la}$, the cavity emission is observed over a range of $\Delta_{ca}$ with peak intensity at a particular value of $\Delta_{ca}$ as shown in Fig. 1(c).
The cavity emission at each $\Delta_{la}$ is maximized by adjusting $\Delta_{ca}$.
Fig. 2(a) shows the maximum intensity achieved at each $\Delta_{la}$ as the total number of incident photons per second.
From Fig. 2(a) we see that maximum cavity emission is observed at $\Delta_{la} \approx -2.5 \Gamma$.
We note that, though the total photon scattering rate from each atom increases considerably as the drive laser is tuned closer to the atomic transition, the observed intensity of the blue detuned emission becomes negligible.
As the drive laser is tuned further off resonance ($\Delta_{la} < -2.5\Gamma$), the cavity emission decreases as the total scattering rate drops.

\begin{figure*}[t]
    \begin{center}
    \includegraphics[width=0.98\linewidth]{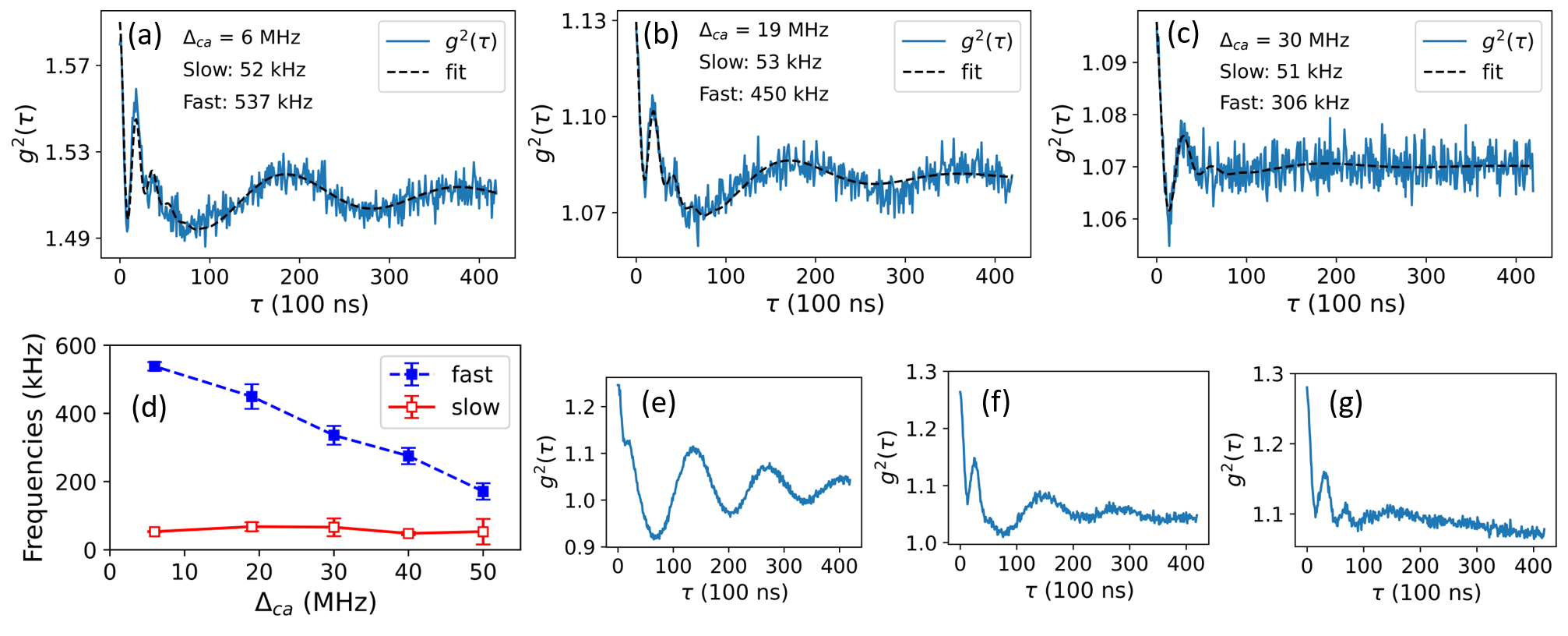}
        \label{Fitting_a_functional_form_on_the_g2}
        \caption{(a-c) show the typical $g^2(\tau)$ measured at $\Delta_{ca}$ = 6, 19 and 30 MHz respectively, with the fit function discussed in the text to obtain the frequencies of the fast and slow oscillations observed in the $g^2(\tau)$. (d) The dependence of the observed frequencies on $\Delta_{ca}$. The solid blue squares and empty red squares indicate the average value of fast and slow frequencies respectively. (e) shows the $g^2(\tau)$ at close to $220^\circ$ polarization angle measured from Z-axis. (f) and (g) show the correlations at $240^\circ$ and $260^\circ$ respectively. Drive laser power is 25 mW, $\Delta_{la} \approx -3\Gamma$, $N_c \approx 23000$, $\Delta_{ca} \approx 10$ MHz.}
	\end{center}
\end{figure*}

Keeping $\Delta_{la}$ constant at $\approx -3 \Gamma$ and power at 25 mW, with $N_c$ $\approx$ 23000, the cavity emission is observed over nearly 80 MHz of $\Delta_{ca}$.
The cavity is locked at various values of $\Delta_{ca}$ in this broad range \cite{SAR_Sreyas} and the rate of photons incident on the single photon detectors is recorded.
Fig. 2(b) shows the total number of incident photons per second.
Depending on the experimental parameters, the signal to noise ratio (SNR) varies from 5:1 to a minimum of 1.5:1.
At values of $\Delta_{ca}$ where SNR is more than 2:1, $g^2(\tau)$ is measured.
The typical graphs of $g^2(\tau)$ at $\Delta_{ca} =$ 6 MHz, 19 MHz and 30 MHz are presented in Fig. 3(a-c). Damped oscillations at two distinct frequencies can be observed in the plots of $g^2(\tau)$. The functional form, $f(\tau) = e^{-\gamma_1\tau}A_1cos(\omega_1\tau + \phi_1) + e^{-\gamma_2\tau} A_2cos(\omega_2\tau + \phi_2) + f_0$ \cite{Rempe_3photon}, is fit to the $g^2(\tau)$.
The average values of faster and slower oscillation frequencies obtained from the fits for each $\Delta_{ca}$ on three sets of $g^2(\tau)$ measurements are shown in Fig. 3(d). The slower frequency is less than 100 kHz and remains range bound for all values of $\Delta_{ca}$, whereas, the faster frequency decreases monotonically from nearly 550 kHz to 170 kHz with increasing $\Delta_{ca}$.
These observations indicate the possible existence of two separate mechanisms for the exchange of energy between the drive field and the field inside the cavity, each with distinct photon statistics.
Interestingly, unlike earlier results \cite{Rempe_3photon}, the frequencies observed in our experiments are not the same as $\Omega_R$ ($\approx$ 23 MHz) or $g_0$ ($\approx$ 200 kHz), which are the frequencies related to atom-photon interactions in our experiment.

For a given longitudinal mode, a cavity can support two mutually independent modes with orthogonal polarizations.
Curious whether the two frequencies seen in the cavity emission exhibit at orthogonal polarizations, we performed another experiment.
A linear polarizer is placed in the cavity output direction along the cavity axis before the multi-mode fiber.
The second order correlations from various polarization components of the blue detuned emission are measured.
The $g^2(\tau)$ functions for a few polarization angles are shown in Fig. 3(e-g).
At a particular polarization angle, the slow oscillations in $g^2(\tau)$ become prominent. Here, the lowest value of $g^2(\tau)$ drops under 1.
At a different polarization, the slow frequency is suppressed and only the fast oscillations in the $g^2(\tau)$ can be observed.
It must be noted that the difference between these two polarization angles which maximize the slow and fast oscillatory behavior is not strictly $\pi/2$.
However, the fact that the two frequencies in $g^2(\tau)$ can largely be isolated in polarization space, confirms two distinct mechanisms for cavity emission by driven MOT atoms.

\begin{figure*}[t]
    \begin{center}
    \includegraphics[width=0.98\linewidth, keepaspectratio]{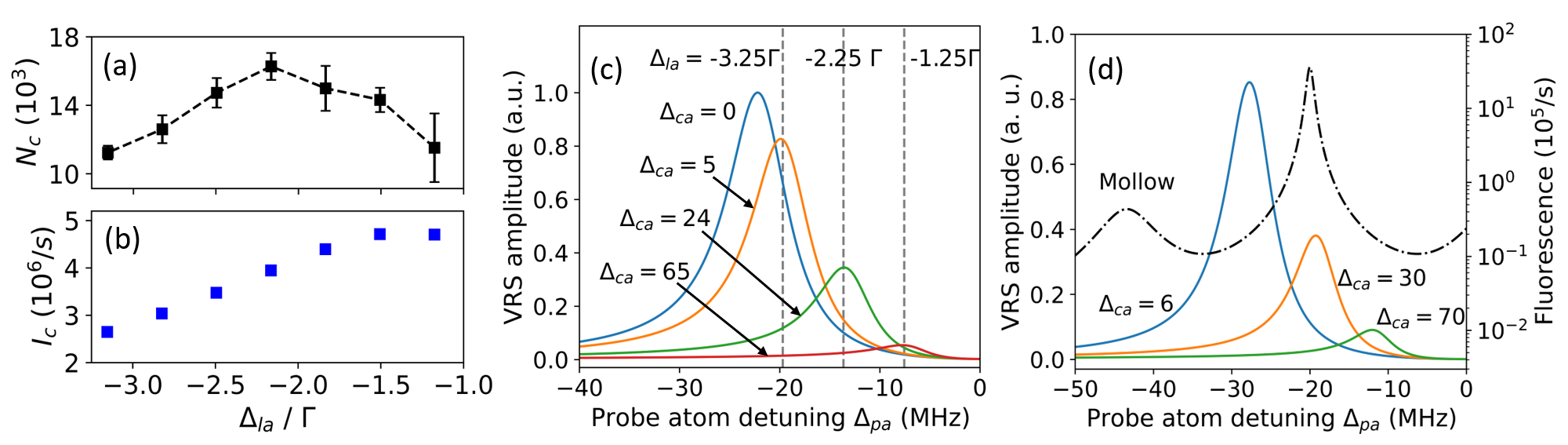}
        \label{Spectral_overlap}
        \caption{All values of $\Delta_{ca}$ are in MHz. (a) The experimentally determined variation in $N_c$ versus $\Delta_{la}$. (b) Calculated variation in coherent photon scattering rate ($I_c$) from each atom versus $\Delta_{la}$. (c) The estimated $\Delta_{ca}$ to match the red side VRS peak with the central Mollow peak ($\omega_l$) for $\Delta_{la} = -1.25\Gamma, -2.25\Gamma$ and $-3.25\Gamma$. The VRS profile depicts the transmission intensity of an arbitrarily weak probe scanning across the atomic transition and incident along the cavity axis. The red VRS amplitudes are normalized to the peak amplitude at $\Delta_{ca} = 0$, curves are labeled for corresponding $\Delta_{ca}$, the dashed lines indicate the relative positions of $\omega_l$ for the three drive detunings. It is shown that the red VRS peak amplitude is considerably small at low $\Delta_{la}$. Drive laser power and $N_c$ are kept constant at 25 mW and 12000 respectively. $N_c$ is chosen close to the experimentally observed value in the $\Delta_{la}$ study. (d) The numerically calculated Mollow spectrum is shown as the black dashed curve. The multi-colour curves are the red VRS profiles at the indicated $\Delta_{ca}$. Maximum cavity intensity is observed at $\Delta_{ca} \approx 30$ MHz where the red VRS profile maximally overlaps the central Mollow peak. For the calculations, drive laser power is 25 mW, $\Delta_{la} = -3\Gamma$, $N_c \approx 23000$.}
	\end{center}
\end{figure*}

\section{Discussion}

To qualitatively understand the dependence of the cavity emission on $\Delta_{la}$ and $\Delta_{ca}$, we consider the spectrum of a driven two level system and the normal modes of $N_c$ two level atoms interacting with a cavity.
The three significant factors that lead to the observed trends in cavity emission are: (a) the rate of coherent scattering per atom, (b) $N_c$ and (c) the fraction of scattered photons accepted by the combined atom-cavity system mode.
Persistent cavity emission implies a coherent buildup of electromagnetic field inside the cavity. 
For 25 mW of drive laser power over a 10 mm beam diameter and $-3.25 \Gamma \leq \Delta_{la} \leq -1.25 \Gamma$, the generalized Rabi frequency ranges from 15 MHz $\leq \Omega_R \leq$ 23 MHz.
Our experimental atom-cavity interaction parameters are $\Gamma$ $\approx$ 6.06 MHz, $\kappa$ $\approx$ 4.4 MHz and $g_0$ $\approx$ 200 kHz.
Since $g_0$ is 2 orders of magnitude smaller than the Rabi frequency of the drive laser, we neglect the effect of individual atom-cavity coupling and therefore, the Mollow spectrum \cite{Mollow_Steck, Scully} of the driven atoms in free space is considered a good approximation.
The coherent scattering rate ($I_c$) can be calculated using standard results for near resonance fluorescence \cite{Mollow_Steck}.
In collective strong coupling regime, the effective strength of interaction between the atoms and the cavity field scales as $g_0 \sqrt{N_c}$ \cite{Agarwal_vrs}.
Since the cooling lasers themselves act as the drive in our experiment, varying the drive laser frequency also varies the MOT atom numbers.
A VRS measurement is performed across the 3-3 transition at each $\Delta_{la}$ to determine $N_c$, which varies between 11000 and 16000 atoms as shown in Fig. 4(a), with a maximum at $\Delta_{la} \approx -2.25 \Gamma$ \cite{SAR_Tridib_Nc, Steck_Rb}.
We define $\Delta_{pa} = (\omega_p - \omega_a) / 2\pi$, where $\omega_p$ is the angular frequency of the probe laser.
The VRS signal of an arbitrarily weak probe scanning across the atomic transition, input along the cavity axis can be calculated using the Tavis-Cummings model \cite{Tavis, Agarwal_vrs, SAR_Rahul}.
Using the coherent scattering rate, the experimentally determined $N_c$ and the calculated VRS peak amplitude, we explain qualitatively, the variation in cavity emission with $\Delta_{la}$ and $\Delta_{ca}$, below.

Tuning the drive laser closer to the atomic transition ($\Delta_{la} \approx -1.25 \Gamma$) decreases the MOT damping force. This results in a 25$\%$ drop in $N_c$ from its maximum value as shown in Fig. 4(a).
At this detuning, the coherent photon scattering rate increases as shown in Fig. 4(b).
At smaller $\Delta_{la}$, the central Mollow peak ($\omega_l$) occurs closer to $\omega_a$.
Therefore, to observe cavity emission, $\Delta_{ca}$ is further blue detuned to shift the red VRS peak towards $\omega_a$ to ensure spectral matching with the central Mollow peak.
However, for a fixed $N_c$, shifting the red VRS peak towards $\omega_a$ decreases its amplitude as shown in Fig. 4(c). Here, the red VRS amplitudes are normalized with respect to its peak amplitude at $\Delta_{ca} = 0$.
At $\Delta_{la} \approx -1.25 \Gamma$ and $N_c \approx 12000$, the spectral matching condition is achieved at $\Delta_{ca} \approx$ 65 MHz.
From Fig. 4(c), the normalized VRS peak amplitude at $\Delta_{ca} \approx$ 65 MHz is approximately 0.05.
This leads to a minimal acceptance of coherently scattered photons into the atom-cavity mode.
Thus, despite the increased coherent scattering, the reduction in $N_c$ and diminished red VRS amplitude result in negligible cavity emission at $\Delta_{la} \approx -1.25 \Gamma$.
At $\Delta_{la} \approx -2.25\Gamma$, the coherent scattering rate reduces marginally while $N_c$ attains its maximum value as shown in Fig. 4(a-b).
Here, at $\Delta_{ca} \approx$ 24 MHz, the red VRS peak aligns with the central Mollow peak with a significantly larger normalized amplitude of 0.35 (Fig. 4(c)).
Thus, higher $N_c$ and significantly increased acceptance rate result in stronger cavity emission intensity at $\Delta_{la} \approx -2.25 \Gamma$.
As the drive laser is tuned further away from the atomic resonance ($\Delta_{la} \approx -3.25\Gamma$), the coherent scattering rate drops by half, and $N_c$ decreases by nearly 30$\%$ of its maximum value, as shown in Fig. 4(a-b).
Therefore, the cavity emission intensity begins to decline.
However, at this detuning, the emission is observed at smaller $\Delta_{ca} (\approx$ 5 MHz) where the normalized red VRS amplitude is even larger (0.8), as shown in Fig. 4(c).
Thus, for $\Delta_{la} \approx -3.25 \Gamma$, despite the overall reduction in intensity, the cavity emission is more as compared to that at $\Delta_{la} \approx -1.25 \Gamma$.

The Mollow spectrum for a drive power of 25 mW at $\Delta_{la} \approx -3 \Gamma$ is shown as the dashed black curve in Fig. 4(d). This figure also shows the variation of the red VRS with $\Delta_{ca}$ at $N_c \approx$ 23000.
From the figure, it can be seen that a spectral overlap exists between the red VRS profile and the central Mollow peak across a wide range of $\Delta_{ca}$.
This overlap explains the nearly 80 MHz width of $\Delta_{ca}$ over which the blue-detuned cavity emission is observed.
The maximum emission intensity occurs when the red VRS peak maximally overlaps with the central Mollow peak.
Thus, this model which treats the near resonant scattering and the VRS separately, effectively captures the observed trends in cavity emission dependence on $\Delta_{la}$ and $\Delta_{ca}$.

To investigate a possible reason for the existence of two oscillations in $g^2(\tau)$, the frequency response of the collectively coupled atoms-cavity system is analysed.
A weak probe laser scanning across the D2 F=3 to F'=4 transition is input along the cavity axis.
The cavity is tuned to the previously identified $\Delta_{ca}$ values.
The intensity of probe transmission along with the blue detuned emission from the cavity is measured on a photo-multiplier tube (PMT).
The probe transmission reveals an additional splitting of the red side VRS peak.
The PMT signal at 6, 19 and 30 MHz of $\Delta_{ca}$ are shown in Fig. 5(a-c).
As seen in the figure, the baselines of the signals recorded in this case are not zero since at these values of $\Delta_{ca}$, there is contribution from both the fluorescing atoms as well as the probe to the light transmitted by the cavity.
As $\Delta_{ca}$ is increased to 30 MHz, the red side VRS peak diminishes while the intensity of emission originating from driven atoms increases and the probe transmission cannot be distinguished from the signal.
The split in the red-side VRS profile is observed at precisely the drive laser frequency.
The origin of this dual-peak structure in the red-side VRS peak is explained by Cheng et al. \cite{Dual_peak_Zhu}, by considering the formation of a lambda-type three-level system involving various $m_F$ levels.
The two frequencies in $g^2(\tau)$ likely originate from the overlap of the two separate parts of the red VRS peak with the central Mollow peak.
The splitting of the red VRS peak highlights the need for a more comprehensive theoretical framework which integrates the near resonant drive field and collective strong coupling.

\begin{figure}[t]
    \begin{center}
    \includegraphics[width=0.47\textwidth, keepaspectratio]{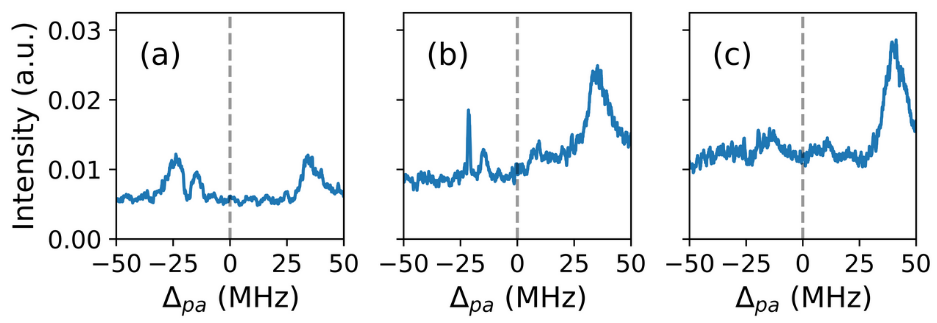}
        \label{Probe_study}
        \caption{Probe transmission at various values of $\Delta_{ca}$. (a), (b) and (c) show the PMT signal for the probe laser transmission with the cavity emission at 6, 19 and 30 MHz values of $\Delta_{ca}$ respectively. The vertical dashed line indicates $\Delta_{pa} = 0$. The plots show two VRS peaks due to collective strong coupling. The non-zero baseline is due to the contribution from blue detuned cavity emission. The splitting in the red side VRS peak is clearly visible at small $\Delta_{ca}$. At larger $\Delta_{ca}$, the red VRS peak diminishes and the cavity emission increases, making the red VRS indistinguishable from the emission signal.}
	\end{center}
\end{figure}

\section{Conclusion}

Much of the prior research on driven atoms interacting with cavities has focused on either the single-atom-cavity strong coupling regime or the bad-cavity limit.
Here we explore the intermediate regime, where a large number of driven atoms collectively interact with the cavity, to understand the statistical properties and underlying mechanisms for cavity emission.
We present experimental observations of the dependence of cavity emission on the drive laser frequency and cavity detuning.
To describe the observed trends, we use a simple model that separately considers the near-resonant drive field and collective atom-cavity interactions.
The measured statistical properties of the cavity emission reveal intriguing effects.
Among the two frequency components in $g^2(\tau)$, one remains largely unaffected by atom-cavity detuning, while the other exhibits a monotonic decrease.
By selecting specific polarization components, it is possible to isolate the frequencies observed in $g^2(\tau)$.
These findings strongly suggest the presence of two distinct, potentially polarization-decoupled mechanisms for energy transfer from the drive field to the cavity field via the atoms.
Analysis based on Tavis-Cummings model does not accurately predict the two oscillation frequencies observed in $g^2(\tau)$.
To address this, it is necessary to develop a theoretical model that can simultaneously treat the effect of the drive field and the collective vacuum Rabi splitting (VRS) response. This model must also account for additional details, such as the polarization of the drive laser and contributions from multiple $m_F$ levels. However, constructing such a model is challenging due to the system's large dimensionality and fluctuations in the total quanta of excitation.
Nevertheless, the study of emission from driven atoms interacting with cavities holds potential for advancing non-destructive detection techniques, achieving ultra-narrow linewidth transitions for atomic clocks and developing cavity-based quantum technologies \cite{Rempe_ND, Reichel_ND, Thompson_g2, SAR_Goku, SAR_Rahul2}.

\section{Acknowledgment}

The authors would like to thank Prof. G. S. Agarwal for his valuable insights and Meena M. S. for her technical support. This research has been funded by the Department of Science and Technology and Ministry of Electronics and Information Technology (MeitY), Government of India, under a Centre for Excellence in Quantum Technologies grant with Ref. No. 4(7)/2020-ITEA.

\bibliography{main}

\begin{thebibliography}{35}%
\makeatletter
\providecommand \@ifxundefined [1]{%
 \@ifx{#1\undefined}
}%
\providecommand \@ifnum [1]{%
 \ifnum #1\expandafter \@firstoftwo
 \else \expandafter \@secondoftwo
 \fi
}%
\providecommand \@ifx [1]{%
 \ifx #1\expandafter \@firstoftwo
 \else \expandafter \@secondoftwo
 \fi
}%
\providecommand \natexlab [1]{#1}%
\providecommand \enquote  [1]{``#1''}%
\providecommand \bibnamefont  [1]{#1}%
\providecommand \bibfnamefont [1]{#1}%
\providecommand \citenamefont [1]{#1}%
\providecommand \href@noop [0]{\@secondoftwo}%
\providecommand \href [0]{\begingroup \@sanitize@url \@href}%
\providecommand \@href[1]{\@@startlink{#1}\@@href}%
\providecommand \@@href[1]{\endgroup#1\@@endlink}%
\providecommand \@sanitize@url [0]{\catcode `\\12\catcode `\$12\catcode
  `\&12\catcode `\#12\catcode `\^12\catcode `\_12\catcode `\%12\relax}%
\providecommand \@@startlink[1]{}%
\providecommand \@@endlink[0]{}%
\providecommand \url  [0]{\begingroup\@sanitize@url \@url }%
\providecommand \@url [1]{\endgroup\@href {#1}{\urlprefix }}%
\providecommand \urlprefix  [0]{URL }%
\providecommand \Eprint [0]{\href }%
\providecommand \doibase [0]{https://doi.org/}%
\providecommand \selectlanguage [0]{\@gobble}%
\providecommand \bibinfo  [0]{\@secondoftwo}%
\providecommand \bibfield  [0]{\@secondoftwo}%
\providecommand \translation [1]{[#1]}%
\providecommand \BibitemOpen [0]{}%
\providecommand \bibitemStop [0]{}%
\providecommand \bibitemNoStop [0]{.\EOS\space}%
\providecommand \EOS [0]{\spacefactor3000\relax}%
\providecommand \BibitemShut  [1]{\csname bibitem#1\endcsname}%
\let\auto@bib@innerbib\@empty
\bibitem [{\citenamefont {Mollow}(1972)}]{Mollow1}%
  \BibitemOpen
  \bibfield  {author} {\bibinfo {author} {\bibfnamefont {B.}~\bibnamefont
  {Mollow}},\ }\href@noop {} {\bibfield  {journal} {\bibinfo  {journal}
  {Physical Review A}\ }\textbf {\bibinfo {volume} {5}},\ \bibinfo {pages}
  {2217} (\bibinfo {year} {1972})}\BibitemShut {NoStop}%
\bibitem [{\citenamefont {Mollow}(1969)}]{Mollow3}%
  \BibitemOpen
  \bibfield  {author} {\bibinfo {author} {\bibfnamefont {B.}~\bibnamefont
  {Mollow}},\ }\href@noop {} {\bibfield  {journal} {\bibinfo  {journal}
  {Physical Review}\ }\textbf {\bibinfo {volume} {188}} (\bibinfo {year}
  {1969})}\BibitemShut {NoStop}%
\bibitem [{\citenamefont {Agarwal}(1998)}]{Mollow2}%
  \BibitemOpen
  \bibfield  {author} {\bibinfo {author} {\bibfnamefont {G.}~\bibnamefont
  {Agarwal}},\ }\href@noop {} {\bibfield  {journal} {\bibinfo  {journal}
  {Journal of Modern Optics}\ }\textbf {\bibinfo {volume} {45}},\ \bibinfo
  {pages} {449} (\bibinfo {year} {1998})}\BibitemShut {NoStop}%
\bibitem [{\citenamefont {Grynberg}\ and\ \citenamefont
  {Cohen-Tannoudji}(1993)}]{Grynberg}%
  \BibitemOpen
  \bibfield  {author} {\bibinfo {author} {\bibfnamefont {G.}~\bibnamefont
  {Grynberg}}\ and\ \bibinfo {author} {\bibfnamefont {C.}~\bibnamefont
  {Cohen-Tannoudji}},\ }\href@noop {} {\bibfield  {journal} {\bibinfo
  {journal} {Optics communications}\ }\textbf {\bibinfo {volume} {96}},\
  \bibinfo {pages} {150} (\bibinfo {year} {1993})}\BibitemShut {NoStop}%
\bibitem [{\citenamefont {Purcell}(1995)}]{purcell}%
  \BibitemOpen
  \bibfield  {author} {\bibinfo {author} {\bibfnamefont {E.~M.}\ \bibnamefont
  {Purcell}},\ }in\ \href@noop {} {\emph {\bibinfo {booktitle} {Confined
  Electrons and Photons: New Physics and Applications}}}\ (\bibinfo
  {publisher} {Springer},\ \bibinfo {year} {1995})\ pp.\ \bibinfo {pages}
  {839--839}\BibitemShut {NoStop}%
\bibitem [{\citenamefont {Kleppner}(1981)}]{Kleppner_Purcell}%
  \BibitemOpen
  \bibfield  {author} {\bibinfo {author} {\bibfnamefont {D.}~\bibnamefont
  {Kleppner}},\ }\href@noop {} {\bibfield  {journal} {\bibinfo  {journal}
  {Physical review letters}\ }\textbf {\bibinfo {volume} {47}},\ \bibinfo
  {pages} {233} (\bibinfo {year} {1981})}\BibitemShut {NoStop}%
\bibitem [{\citenamefont {Lezama}\ \emph {et~al.}(1990)\citenamefont {Lezama},
  \citenamefont {Zhu}, \citenamefont {Kanskar},\ and\ \citenamefont
  {Mossberg}}]{Mossberg}%
  \BibitemOpen
  \bibfield  {author} {\bibinfo {author} {\bibfnamefont {A.}~\bibnamefont
  {Lezama}}, \bibinfo {author} {\bibfnamefont {Y.}~\bibnamefont {Zhu}},
  \bibinfo {author} {\bibfnamefont {M.}~\bibnamefont {Kanskar}},\ and\ \bibinfo
  {author} {\bibfnamefont {T.}~\bibnamefont {Mossberg}},\ }\href@noop {}
  {\bibfield  {journal} {\bibinfo  {journal} {Physical Review A}\ }\textbf
  {\bibinfo {volume} {41}},\ \bibinfo {pages} {1576} (\bibinfo {year}
  {1990})}\BibitemShut {NoStop}%
\bibitem [{\citenamefont {Jaynes}\ and\ \citenamefont
  {Cummings}(1963)}]{Jaynes_Cummings_SC}%
  \BibitemOpen
  \bibfield  {author} {\bibinfo {author} {\bibfnamefont {E.~T.}\ \bibnamefont
  {Jaynes}}\ and\ \bibinfo {author} {\bibfnamefont {F.~W.}\ \bibnamefont
  {Cummings}},\ }\href@noop {} {\bibfield  {journal} {\bibinfo  {journal}
  {Proceedings of the IEEE}\ }\textbf {\bibinfo {volume} {51}},\ \bibinfo
  {pages} {89} (\bibinfo {year} {1963})}\BibitemShut {NoStop}%
\bibitem [{\citenamefont {Thompson}\ \emph {et~al.}(1992)\citenamefont
  {Thompson}, \citenamefont {Rempe},\ and\ \citenamefont {Kimble}}]{Rempe_VRS}%
  \BibitemOpen
  \bibfield  {author} {\bibinfo {author} {\bibfnamefont {R.}~\bibnamefont
  {Thompson}}, \bibinfo {author} {\bibfnamefont {G.}~\bibnamefont {Rempe}},\
  and\ \bibinfo {author} {\bibfnamefont {H.}~\bibnamefont {Kimble}},\
  }\href@noop {} {\bibfield  {journal} {\bibinfo  {journal} {Physical review
  letters}\ }\textbf {\bibinfo {volume} {68}},\ \bibinfo {pages} {1132}
  (\bibinfo {year} {1992})}\BibitemShut {NoStop}%
\bibitem [{\citenamefont {Colombe}\ \emph {et~al.}(2007)\citenamefont
  {Colombe}, \citenamefont {Steinmetz}, \citenamefont {Dubois}, \citenamefont
  {Linke}, \citenamefont {Hunger},\ and\ \citenamefont {Reichel}}]{BEC_cav}%
  \BibitemOpen
  \bibfield  {author} {\bibinfo {author} {\bibfnamefont {Y.}~\bibnamefont
  {Colombe}}, \bibinfo {author} {\bibfnamefont {T.}~\bibnamefont {Steinmetz}},
  \bibinfo {author} {\bibfnamefont {G.}~\bibnamefont {Dubois}}, \bibinfo
  {author} {\bibfnamefont {F.}~\bibnamefont {Linke}}, \bibinfo {author}
  {\bibfnamefont {D.}~\bibnamefont {Hunger}},\ and\ \bibinfo {author}
  {\bibfnamefont {J.}~\bibnamefont {Reichel}},\ }\href@noop {} {\bibfield
  {journal} {\bibinfo  {journal} {Nature}\ }\textbf {\bibinfo {volume} {450}},\
  \bibinfo {pages} {272} (\bibinfo {year} {2007})}\BibitemShut {NoStop}%
\bibitem [{\citenamefont {Vrijsen}\ \emph {et~al.}(2011)\citenamefont
  {Vrijsen}, \citenamefont {Hosten}, \citenamefont {Lee}, \citenamefont
  {Bernon},\ and\ \citenamefont {Kasevich}}]{Kasevich}%
  \BibitemOpen
  \bibfield  {author} {\bibinfo {author} {\bibfnamefont {G.}~\bibnamefont
  {Vrijsen}}, \bibinfo {author} {\bibfnamefont {O.}~\bibnamefont {Hosten}},
  \bibinfo {author} {\bibfnamefont {J.}~\bibnamefont {Lee}}, \bibinfo {author}
  {\bibfnamefont {S.}~\bibnamefont {Bernon}},\ and\ \bibinfo {author}
  {\bibfnamefont {M.~A.}\ \bibnamefont {Kasevich}},\ }\href@noop {} {\bibfield
  {journal} {\bibinfo  {journal} {Physical review letters}\ }\textbf {\bibinfo
  {volume} {107}},\ \bibinfo {pages} {063904} (\bibinfo {year}
  {2011})}\BibitemShut {NoStop}%
\bibitem [{\citenamefont {Raizen}\ \emph {et~al.}(1989)\citenamefont {Raizen},
  \citenamefont {Thompson}, \citenamefont {Brecha}, \citenamefont {Kimble},\
  and\ \citenamefont {Carmichael}}]{Brecha_Carmichael_1}%
  \BibitemOpen
  \bibfield  {author} {\bibinfo {author} {\bibfnamefont {M.}~\bibnamefont
  {Raizen}}, \bibinfo {author} {\bibfnamefont {R.}~\bibnamefont {Thompson}},
  \bibinfo {author} {\bibfnamefont {R.}~\bibnamefont {Brecha}}, \bibinfo
  {author} {\bibfnamefont {H.}~\bibnamefont {Kimble}},\ and\ \bibinfo {author}
  {\bibfnamefont {H.}~\bibnamefont {Carmichael}},\ }\href@noop {} {\bibfield
  {journal} {\bibinfo  {journal} {Physical review letters}\ }\textbf {\bibinfo
  {volume} {63}},\ \bibinfo {pages} {240} (\bibinfo {year} {1989})}\BibitemShut
  {NoStop}%
\bibitem [{\citenamefont {Carmichael}(1986)}]{carmichael}%
  \BibitemOpen
  \bibfield  {author} {\bibinfo {author} {\bibfnamefont {H.}~\bibnamefont
  {Carmichael}},\ }\href@noop {} {\bibfield  {journal} {\bibinfo  {journal}
  {Physical Review A}\ }\textbf {\bibinfo {volume} {33}},\ \bibinfo {pages}
  {3262} (\bibinfo {year} {1986})}\BibitemShut {NoStop}%
\bibitem [{\citenamefont {Zhou}\ and\ \citenamefont
  {Swain}(1998)}]{Zhou_Swain}%
  \BibitemOpen
  \bibfield  {author} {\bibinfo {author} {\bibfnamefont {P.}~\bibnamefont
  {Zhou}}\ and\ \bibinfo {author} {\bibfnamefont {S.}~\bibnamefont {Swain}},\
  }\href@noop {} {\bibfield  {journal} {\bibinfo  {journal} {Physical Review
  A}\ }\textbf {\bibinfo {volume} {58}},\ \bibinfo {pages} {1515} (\bibinfo
  {year} {1998})}\BibitemShut {NoStop}%
\bibitem [{\citenamefont {Tavis}\ and\ \citenamefont {Cummings}(1968)}]{Tavis}%
  \BibitemOpen
  \bibfield  {author} {\bibinfo {author} {\bibfnamefont {M.}~\bibnamefont
  {Tavis}}\ and\ \bibinfo {author} {\bibfnamefont {F.~W.}\ \bibnamefont
  {Cummings}},\ }\href@noop {} {\bibfield  {journal} {\bibinfo  {journal}
  {Physical Review}\ }\textbf {\bibinfo {volume} {170}},\ \bibinfo {pages}
  {379} (\bibinfo {year} {1968})}\BibitemShut {NoStop}%
\bibitem [{\citenamefont {Agarwal}(1984)}]{Agarwal_vrs}%
  \BibitemOpen
  \bibfield  {author} {\bibinfo {author} {\bibfnamefont {G.}~\bibnamefont
  {Agarwal}},\ }\href@noop {} {\bibfield  {journal} {\bibinfo  {journal}
  {Physical review letters}\ }\textbf {\bibinfo {volume} {53}},\ \bibinfo
  {pages} {1732} (\bibinfo {year} {1984})}\BibitemShut {NoStop}%
\bibitem [{\citenamefont {Hernandez}\ \emph {et~al.}(2009)\citenamefont
  {Hernandez}, \citenamefont {Zhang},\ and\ \citenamefont {Zhu}}]{Zhu_Line}%
  \BibitemOpen
  \bibfield  {author} {\bibinfo {author} {\bibfnamefont {G.}~\bibnamefont
  {Hernandez}}, \bibinfo {author} {\bibfnamefont {J.}~\bibnamefont {Zhang}},\
  and\ \bibinfo {author} {\bibfnamefont {Y.}~\bibnamefont {Zhu}},\ }\href@noop
  {} {\bibfield  {journal} {\bibinfo  {journal} {Optics Express}\ }\textbf
  {\bibinfo {volume} {17}},\ \bibinfo {pages} {4798} (\bibinfo {year}
  {2009})}\BibitemShut {NoStop}%
\bibitem [{\citenamefont {Ray}\ \emph {et~al.}(2013)\citenamefont {Ray},
  \citenamefont {Sharma}, \citenamefont {Jyothi},\ and\ \citenamefont
  {Rangwala}}]{SAR_Tridib_Nc}%
  \BibitemOpen
  \bibfield  {author} {\bibinfo {author} {\bibfnamefont {T.}~\bibnamefont
  {Ray}}, \bibinfo {author} {\bibfnamefont {A.}~\bibnamefont {Sharma}},
  \bibinfo {author} {\bibfnamefont {S.}~\bibnamefont {Jyothi}},\ and\ \bibinfo
  {author} {\bibfnamefont {S.}~\bibnamefont {Rangwala}},\ }\href@noop {}
  {\bibfield  {journal} {\bibinfo  {journal} {Physical Review A}\ }\textbf
  {\bibinfo {volume} {87}},\ \bibinfo {pages} {033832} (\bibinfo {year}
  {2013})}\BibitemShut {NoStop}%
\bibitem [{\citenamefont {Dutta}\ and\ \citenamefont {Rangwala}(2016)}]{SAR4}%
  \BibitemOpen
  \bibfield  {author} {\bibinfo {author} {\bibfnamefont {S.}~\bibnamefont
  {Dutta}}\ and\ \bibinfo {author} {\bibfnamefont {S.}~\bibnamefont
  {Rangwala}},\ }\href@noop {} {\bibfield  {journal} {\bibinfo  {journal}
  {Physical Review A}\ }\textbf {\bibinfo {volume} {94}},\ \bibinfo {pages}
  {053841} (\bibinfo {year} {2016})}\BibitemShut {NoStop}%
\bibitem [{\citenamefont {Gokul}\ \emph {et~al.}(2024)\citenamefont {Gokul},
  \citenamefont {Bahuleyan}, \citenamefont {Dinesh}, \citenamefont {Thakar},\
  and\ \citenamefont {Rangwala}}]{SAR_Goku}%
  \BibitemOpen
  \bibfield  {author} {\bibinfo {author} {\bibfnamefont {V.}~\bibnamefont
  {Gokul}}, \bibinfo {author} {\bibfnamefont {A.}~\bibnamefont {Bahuleyan}},
  \bibinfo {author} {\bibfnamefont {S.}~\bibnamefont {Dinesh}}, \bibinfo
  {author} {\bibfnamefont {V.}~\bibnamefont {Thakar}},\ and\ \bibinfo {author}
  {\bibfnamefont {S.}~\bibnamefont {Rangwala}},\ }\href@noop {} {\bibfield
  {journal} {\bibinfo  {journal} {Physical Review A}\ }\textbf {\bibinfo
  {volume} {109}},\ \bibinfo {pages} {053713} (\bibinfo {year}
  {2024})}\BibitemShut {NoStop}%
\bibitem [{\citenamefont {Sawant}\ and\ \citenamefont
  {Rangwala}(2017)}]{SAR_Rahul}%
  \BibitemOpen
  \bibfield  {author} {\bibinfo {author} {\bibfnamefont {R.}~\bibnamefont
  {Sawant}}\ and\ \bibinfo {author} {\bibfnamefont {S.}~\bibnamefont
  {Rangwala}},\ }\href@noop {} {\bibfield  {journal} {\bibinfo  {journal}
  {Scientific Reports}\ }\textbf {\bibinfo {volume} {7}},\ \bibinfo {pages}
  {11432} (\bibinfo {year} {2017})}\BibitemShut {NoStop}%
\bibitem [{\citenamefont {Bahuleyan}\ \emph {et~al.}(2025)\citenamefont
  {Bahuleyan}, \citenamefont {Thakar}, \citenamefont {Gokul}, \citenamefont
  {Dinesh}, \citenamefont {Prasanna~Venkatesh},\ and\ \citenamefont
  {SA}}]{SAR_Arun}%
  \BibitemOpen
  \bibfield  {author} {\bibinfo {author} {\bibfnamefont {A.}~\bibnamefont
  {Bahuleyan}}, \bibinfo {author} {\bibfnamefont {V.}~\bibnamefont {Thakar}},
  \bibinfo {author} {\bibfnamefont {V.}~\bibnamefont {Gokul}}, \bibinfo
  {author} {\bibfnamefont {S.}~\bibnamefont {Dinesh}}, \bibinfo {author}
  {\bibfnamefont {B.}~\bibnamefont {Prasanna~Venkatesh}},\ and\ \bibinfo
  {author} {\bibfnamefont {R.}~\bibnamefont {SA}}} (\bibinfo {year} {2025}),\
  \bibinfo {note} {gain, amplification and lasing in driven atom-cavity system
  (Unpublished manuscript)}\BibitemShut {NoStop}%
\bibitem [{\citenamefont {Hanbury~Brown}\ and\ \citenamefont
  {Twiss}(1979)}]{HBT}%
  \BibitemOpen
  \bibfield  {author} {\bibinfo {author} {\bibfnamefont {R.}~\bibnamefont
  {Hanbury~Brown}}\ and\ \bibinfo {author} {\bibfnamefont {R.~Q.}\ \bibnamefont
  {Twiss}},\ }in\ \href@noop {} {\emph {\bibinfo {booktitle} {A Source Book in
  Astronomy and Astrophysics, 1900--1975}}}\ (\bibinfo  {publisher} {Harvard
  University Press},\ \bibinfo {year} {1979})\ pp.\ \bibinfo {pages}
  {8--12}\BibitemShut {NoStop}%
\bibitem [{\citenamefont {Shafi}\ \emph {et~al.}(2015)\citenamefont {Shafi},
  \citenamefont {Pandey}, \citenamefont {Suryabrahmam}, \citenamefont
  {Girish},\ and\ \citenamefont {Ramachandran}}]{Shafi}%
  \BibitemOpen
  \bibfield  {author} {\bibinfo {author} {\bibfnamefont {K.~M.}\ \bibnamefont
  {Shafi}}, \bibinfo {author} {\bibfnamefont {D.}~\bibnamefont {Pandey}},
  \bibinfo {author} {\bibfnamefont {B.}~\bibnamefont {Suryabrahmam}}, \bibinfo
  {author} {\bibfnamefont {B.}~\bibnamefont {Girish}},\ and\ \bibinfo {author}
  {\bibfnamefont {H.}~\bibnamefont {Ramachandran}},\ }\href@noop {} {\bibfield
  {journal} {\bibinfo  {journal} {Journal of Physics B: Atomic, Molecular and
  Optical Physics}\ }\textbf {\bibinfo {volume} {49}},\ \bibinfo {pages}
  {025301} (\bibinfo {year} {2015})}\BibitemShut {NoStop}%
\bibitem [{\citenamefont {Laurence}\ \emph {et~al.}(2006)\citenamefont
  {Laurence}, \citenamefont {Fore},\ and\ \citenamefont {Huser}}]{Python}%
  \BibitemOpen
  \bibfield  {author} {\bibinfo {author} {\bibfnamefont {T.~A.}\ \bibnamefont
  {Laurence}}, \bibinfo {author} {\bibfnamefont {S.}~\bibnamefont {Fore}},\
  and\ \bibinfo {author} {\bibfnamefont {T.}~\bibnamefont {Huser}},\
  }\href@noop {} {\bibfield  {journal} {\bibinfo  {journal} {Optics letters}\
  }\textbf {\bibinfo {volume} {31}},\ \bibinfo {pages} {829} (\bibinfo {year}
  {2006})}\BibitemShut {NoStop}%
\bibitem [{\citenamefont {Dinesh}\ \emph {et~al.}(2024)\citenamefont {Dinesh},
  \citenamefont {Thakar}, \citenamefont {Gokul}, \citenamefont {Bahuleyan},\
  and\ \citenamefont {Rangwala}}]{SAR_Sreyas}%
  \BibitemOpen
  \bibfield  {author} {\bibinfo {author} {\bibfnamefont {S.}~\bibnamefont
  {Dinesh}}, \bibinfo {author} {\bibfnamefont {V.}~\bibnamefont {Thakar}},
  \bibinfo {author} {\bibfnamefont {V.}~\bibnamefont {Gokul}}, \bibinfo
  {author} {\bibfnamefont {A.}~\bibnamefont {Bahuleyan}},\ and\ \bibinfo
  {author} {\bibfnamefont {S.}~\bibnamefont {Rangwala}},\ }\href@noop {}
  {\bibfield  {journal} {\bibinfo  {journal} {EPJ Techniques and
  Instrumentation}\ }\textbf {\bibinfo {volume} {11}},\ \bibinfo {pages} {1}
  (\bibinfo {year} {2024})}\BibitemShut {NoStop}%
\bibitem [{\citenamefont {Koch}\ \emph {et~al.}(2011)\citenamefont {Koch},
  \citenamefont {Sames}, \citenamefont {Balbach}, \citenamefont {Chibani},
  \citenamefont {Kubanek}, \citenamefont {Murr}, \citenamefont {Wilk},\ and\
  \citenamefont {Rempe}}]{Rempe_3photon}%
  \BibitemOpen
  \bibfield  {author} {\bibinfo {author} {\bibfnamefont {M.}~\bibnamefont
  {Koch}}, \bibinfo {author} {\bibfnamefont {C.}~\bibnamefont {Sames}},
  \bibinfo {author} {\bibfnamefont {M.}~\bibnamefont {Balbach}}, \bibinfo
  {author} {\bibfnamefont {H.}~\bibnamefont {Chibani}}, \bibinfo {author}
  {\bibfnamefont {A.}~\bibnamefont {Kubanek}}, \bibinfo {author} {\bibfnamefont
  {K.}~\bibnamefont {Murr}}, \bibinfo {author} {\bibfnamefont {T.}~\bibnamefont
  {Wilk}},\ and\ \bibinfo {author} {\bibfnamefont {G.}~\bibnamefont {Rempe}},\
  }\href@noop {} {\bibfield  {journal} {\bibinfo  {journal} {Physical review
  letters}\ }\textbf {\bibinfo {volume} {107}},\ \bibinfo {pages} {023601}
  (\bibinfo {year} {2011})}\BibitemShut {NoStop}%
\bibitem [{\citenamefont {Steck}(2011)}]{Mollow_Steck}%
  \BibitemOpen
  \bibfield  {author} {\bibinfo {author} {\bibfnamefont {D.~A.}\ \bibnamefont
  {Steck}},\ }\href@noop {} {\bibinfo {title} {Quantum and atom optics}}
  (\bibinfo {year} {2011})\BibitemShut {NoStop}%
\bibitem [{\citenamefont {Scully}\ and\ \citenamefont
  {Zubairy}(1997)}]{Scully}%
  \BibitemOpen
  \bibfield  {author} {\bibinfo {author} {\bibfnamefont {M.~O.}\ \bibnamefont
  {Scully}}\ and\ \bibinfo {author} {\bibfnamefont {M.~S.}\ \bibnamefont
  {Zubairy}},\ }\href@noop {} {\emph {\bibinfo {title} {Quantum optics}}}\
  (\bibinfo  {publisher} {Cambridge university press},\ \bibinfo {year}
  {1997})\BibitemShut {NoStop}%
\bibitem [{\citenamefont {Steck}(2013)}]{Steck_Rb}%
  \BibitemOpen
  \bibfield  {author} {\bibinfo {author} {\bibfnamefont {D.~A.}\ \bibnamefont
  {Steck}},\ }\href@noop {} {\bibinfo {title} {Rubidium 85 d line data,
  september}} (\bibinfo {year} {2013})\BibitemShut {NoStop}%
\bibitem [{\citenamefont {Cheng}\ \emph {et~al.}(2016)\citenamefont {Cheng},
  \citenamefont {Tan}, \citenamefont {Wang}, \citenamefont {Zhu},\ and\
  \citenamefont {Zhan}}]{Dual_peak_Zhu}%
  \BibitemOpen
  \bibfield  {author} {\bibinfo {author} {\bibfnamefont {Y.}~\bibnamefont
  {Cheng}}, \bibinfo {author} {\bibfnamefont {Z.}~\bibnamefont {Tan}}, \bibinfo
  {author} {\bibfnamefont {J.}~\bibnamefont {Wang}}, \bibinfo {author}
  {\bibfnamefont {Y.-F.}\ \bibnamefont {Zhu}},\ and\ \bibinfo {author}
  {\bibfnamefont {M.-S.}\ \bibnamefont {Zhan}},\ }\href@noop {} {\bibfield
  {journal} {\bibinfo  {journal} {Chinese Physics Letters}\ }\textbf {\bibinfo
  {volume} {33}},\ \bibinfo {pages} {014202} (\bibinfo {year}
  {2016})}\BibitemShut {NoStop}%
\bibitem [{\citenamefont {Bochmann}\ \emph {et~al.}(2010)\citenamefont
  {Bochmann}, \citenamefont {M{\"u}cke}, \citenamefont {Guhl}, \citenamefont
  {Ritter}, \citenamefont {Rempe},\ and\ \citenamefont {Moehring}}]{Rempe_ND}%
  \BibitemOpen
  \bibfield  {author} {\bibinfo {author} {\bibfnamefont {J.}~\bibnamefont
  {Bochmann}}, \bibinfo {author} {\bibfnamefont {M.}~\bibnamefont {M{\"u}cke}},
  \bibinfo {author} {\bibfnamefont {C.}~\bibnamefont {Guhl}}, \bibinfo {author}
  {\bibfnamefont {S.}~\bibnamefont {Ritter}}, \bibinfo {author} {\bibfnamefont
  {G.}~\bibnamefont {Rempe}},\ and\ \bibinfo {author} {\bibfnamefont {D.~L.}\
  \bibnamefont {Moehring}},\ }\href@noop {} {\bibfield  {journal} {\bibinfo
  {journal} {Physical review letters}\ }\textbf {\bibinfo {volume} {104}},\
  \bibinfo {pages} {203601} (\bibinfo {year} {2010})}\BibitemShut {NoStop}%
\bibitem [{\citenamefont {Gehr}\ \emph {et~al.}(2010)\citenamefont {Gehr},
  \citenamefont {Volz}, \citenamefont {Dubois}, \citenamefont {Steinmetz},
  \citenamefont {Colombe}, \citenamefont {Lev}, \citenamefont {Long},
  \citenamefont {Esteve},\ and\ \citenamefont {Reichel}}]{Reichel_ND}%
  \BibitemOpen
  \bibfield  {author} {\bibinfo {author} {\bibfnamefont {R.}~\bibnamefont
  {Gehr}}, \bibinfo {author} {\bibfnamefont {J.}~\bibnamefont {Volz}}, \bibinfo
  {author} {\bibfnamefont {G.}~\bibnamefont {Dubois}}, \bibinfo {author}
  {\bibfnamefont {T.}~\bibnamefont {Steinmetz}}, \bibinfo {author}
  {\bibfnamefont {Y.}~\bibnamefont {Colombe}}, \bibinfo {author} {\bibfnamefont
  {B.~L.}\ \bibnamefont {Lev}}, \bibinfo {author} {\bibfnamefont
  {R.}~\bibnamefont {Long}}, \bibinfo {author} {\bibfnamefont {J.}~\bibnamefont
  {Esteve}},\ and\ \bibinfo {author} {\bibfnamefont {J.}~\bibnamefont
  {Reichel}},\ }\href@noop {} {\bibfield  {journal} {\bibinfo  {journal}
  {Physical review letters}\ }\textbf {\bibinfo {volume} {104}},\ \bibinfo
  {pages} {203602} (\bibinfo {year} {2010})}\BibitemShut {NoStop}%
\bibitem [{\citenamefont {Sch{\"a}fer}\ \emph {et~al.}(2024)\citenamefont
  {Sch{\"a}fer}, \citenamefont {Niu}, \citenamefont {Cline}, \citenamefont
  {Young}, \citenamefont {Song}, \citenamefont {Ritsch},\ and\ \citenamefont
  {Thompson}}]{Thompson_g2}%
  \BibitemOpen
  \bibfield  {author} {\bibinfo {author} {\bibfnamefont {V.}~\bibnamefont
  {Sch{\"a}fer}}, \bibinfo {author} {\bibfnamefont {Z.}~\bibnamefont {Niu}},
  \bibinfo {author} {\bibfnamefont {J.}~\bibnamefont {Cline}}, \bibinfo
  {author} {\bibfnamefont {D.}~\bibnamefont {Young}}, \bibinfo {author}
  {\bibfnamefont {E.}~\bibnamefont {Song}}, \bibinfo {author} {\bibfnamefont
  {H.}~\bibnamefont {Ritsch}},\ and\ \bibinfo {author} {\bibfnamefont
  {J.}~\bibnamefont {Thompson}},\ }\href@noop {} {\bibfield  {journal}
  {\bibinfo  {journal} {arXiv preprint arXiv:2405.20952}\ } (\bibinfo {year}
  {2024})}\BibitemShut {NoStop}%
\bibitem [{\citenamefont {Sawant}\ \emph {et~al.}(2018)\citenamefont {Sawant},
  \citenamefont {Dulieu},\ and\ \citenamefont {Rangwala}}]{SAR_Rahul2}%
  \BibitemOpen
  \bibfield  {author} {\bibinfo {author} {\bibfnamefont {R.}~\bibnamefont
  {Sawant}}, \bibinfo {author} {\bibfnamefont {O.}~\bibnamefont {Dulieu}},\
  and\ \bibinfo {author} {\bibfnamefont {S.~A.}\ \bibnamefont {Rangwala}},\
  }\href@noop {} {\bibfield  {journal} {\bibinfo  {journal} {Physical Review
  A}\ }\textbf {\bibinfo {volume} {97}},\ \bibinfo {pages} {063405} (\bibinfo
  {year} {2018})}\BibitemShut {NoStop}%
\end{thebibliography}%


\begin{thebibliography}{6}%
\makeatletter
\providecommand \@ifxundefined [1]{%
 \@ifx{#1\undefined}
}%
\providecommand \@ifnum [1]{%
 \ifnum #1\expandafter \@firstoftwo
 \else \expandafter \@secondoftwo
 \fi
}%
\providecommand \@ifx [1]{%
 \ifx #1\expandafter \@firstoftwo
 \else \expandafter \@secondoftwo
 \fi
}%
\providecommand \natexlab [1]{#1}%
\providecommand \enquote  [1]{``#1''}%
\providecommand \bibnamefont  [1]{#1}%
\providecommand \bibfnamefont [1]{#1}%
\providecommand \citenamefont [1]{#1}%
\providecommand \href@noop [0]{\@secondoftwo}%
\providecommand \href [0]{\begingroup \@sanitize@url \@href}%
\providecommand \@href[1]{\@@startlink{#1}\@@href}%
\providecommand \@@href[1]{\endgroup#1\@@endlink}%
\providecommand \@sanitize@url [0]{\catcode `\\12\catcode `\$12\catcode
  `\&12\catcode `\#12\catcode `\^12\catcode `\_12\catcode `\%12\relax}%
\providecommand \@@startlink[1]{}%
\providecommand \@@endlink[0]{}%
\providecommand \url  [0]{\begingroup\@sanitize@url \@url }%
\providecommand \@url [1]{\endgroup\@href {#1}{\urlprefix }}%
\providecommand \urlprefix  [0]{URL }%
\providecommand \Eprint [0]{\href }%
\providecommand \doibase [0]{https://doi.org/}%
\providecommand \selectlanguage [0]{\@gobble}%
\providecommand \bibinfo  [0]{\@secondoftwo}%
\providecommand \bibfield  [0]{\@secondoftwo}%
\providecommand \translation [1]{[#1]}%
\providecommand \BibitemOpen [0]{}%
\providecommand \bibitemStop [0]{}%
\providecommand \bibitemNoStop [0]{.\EOS\space}%
\providecommand \EOS [0]{\spacefactor3000\relax}%
\providecommand \BibitemShut  [1]{\csname bibitem#1\endcsname}%
\let\auto@bib@innerbib\@empty
\bibitem [{\citenamefont {Steck}(2011)}]{Mollow_Steck}%
  \BibitemOpen
  \bibfield  {author} {\bibinfo {author} {\bibfnamefont {D.~A.}\ \bibnamefont
  {Steck}},\ }\href@noop {} {\bibinfo {title} {Quantum and atom optics}}
  (\bibinfo {year} {2011})\BibitemShut {NoStop}%
\bibitem [{\citenamefont {Mollow}(1972)}]{Mollow1}%
  \BibitemOpen
  \bibfield  {author} {\bibinfo {author} {\bibfnamefont {B.}~\bibnamefont
  {Mollow}},\ }\href@noop {} {\bibfield  {journal} {\bibinfo  {journal}
  {Physical Review A}\ }\textbf {\bibinfo {volume} {5}},\ \bibinfo {pages}
  {2217} (\bibinfo {year} {1972})}\BibitemShut {NoStop}%
\bibitem [{\citenamefont {Carmichael}(2009)}]{carmichael_lecture}%
  \BibitemOpen
  \bibfield  {author} {\bibinfo {author} {\bibfnamefont {H.}~\bibnamefont
  {Carmichael}},\ }\href@noop {} {\emph {\bibinfo {title} {An open systems
  approach to quantum optics: lectures presented at the Universit{\'e} Libre de
  Bruxelles, October 28 to November 4, 1991}}},\ Vol.~\bibinfo {volume} {18}\
  (\bibinfo  {publisher} {Springer Science \& Business Media},\ \bibinfo {year}
  {2009})\BibitemShut {NoStop}%
\bibitem [{\citenamefont {Sawant}\ and\ \citenamefont
  {Rangwala}(2017)}]{SAR_Rahul}%
  \BibitemOpen
  \bibfield  {author} {\bibinfo {author} {\bibfnamefont {R.}~\bibnamefont
  {Sawant}}\ and\ \bibinfo {author} {\bibfnamefont {S.}~\bibnamefont
  {Rangwala}},\ }\href@noop {} {\bibfield  {journal} {\bibinfo  {journal}
  {Scientific Reports}\ }\textbf {\bibinfo {volume} {7}},\ \bibinfo {pages}
  {11432} (\bibinfo {year} {2017})}\BibitemShut {NoStop}%
\bibitem [{\citenamefont {Sawant}(2018)}]{rahul_thesis}%
  \BibitemOpen
  \bibfield  {author} {\bibinfo {author} {\bibfnamefont {R.}~\bibnamefont
  {Sawant}},\ }\emph {\bibinfo {title} {Interactions between ultracold dilute
  gas of atoms, ions, and cavity}},\ \href@noop {} {Ph.D. thesis},\ \bibinfo
  {school} {Raman Research Institute, Bangalore.} (\bibinfo {year}
  {2018})\BibitemShut {NoStop}%
\bibitem [{\citenamefont {Koch}\ \emph {et~al.}(2011)\citenamefont {Koch},
  \citenamefont {Sames}, \citenamefont {Balbach}, \citenamefont {Chibani},
  \citenamefont {Kubanek}, \citenamefont {Murr}, \citenamefont {Wilk},\ and\
  \citenamefont {Rempe}}]{Rempe_3photon}%
  \BibitemOpen
  \bibfield  {author} {\bibinfo {author} {\bibfnamefont {M.}~\bibnamefont
  {Koch}}, \bibinfo {author} {\bibfnamefont {C.}~\bibnamefont {Sames}},
  \bibinfo {author} {\bibfnamefont {M.}~\bibnamefont {Balbach}}, \bibinfo
  {author} {\bibfnamefont {H.}~\bibnamefont {Chibani}}, \bibinfo {author}
  {\bibfnamefont {A.}~\bibnamefont {Kubanek}}, \bibinfo {author} {\bibfnamefont
  {K.}~\bibnamefont {Murr}}, \bibinfo {author} {\bibfnamefont {T.}~\bibnamefont
  {Wilk}},\ and\ \bibinfo {author} {\bibfnamefont {G.}~\bibnamefont {Rempe}},\
  }\href@noop {} {\bibfield  {journal} {\bibinfo  {journal} {Physical review
  letters}\ }\textbf {\bibinfo {volume} {107}},\ \bibinfo {pages} {023601}
  (\bibinfo {year} {2011})}\BibitemShut {NoStop}%
\end{thebibliography}%

\end{document}


\title{Supplemental Material: Emission from driven atoms in collective strong coupling with an optical cavity }
 \author{V.R. Thakar}%
 \email{vardhanr@rri.res.in}

 \author{A. Bahuleyan}%

 \author{V. I. Gokul}%

 \author{S. P. Dinesh}%

 \author{S. A. Rangwala}%
  \email{sarangwala@rri.res.in}

 \affiliation{%
  Raman Research Institute, C. V. Raman Avenue, Sadashivanagar, Bangalore 560080, India \\}

\maketitle

\section{Near resonance fluorescence calculation}

Let the state of the two level atom be $\ket{\psi}=c_g\ket{g} + c_e\ket{e}$, where we label the ground and excited states of the atom as $\ket{g}$ and $\ket{e}$ respectively.
We define the drive laser detuning $\Delta_{la} = (\omega_l-\omega_a)/$, where, $\omega_l$ and $\omega_a$ are the angular frequencies of the drive laser and the atomic transition respectively.
We define the atomic raising and lower operators as $\sigma^\dagger \mathrel{:=}\ket{e}\bra{g}; \sigma \mathrel{:=} \ket{g}\bra{e}$.
Let the electric field amplitude of the drive laser be $E_0$, then the dipole interaction between the drive field and the atom is characterized by the Rabi frequency given by $\Omega_r \mathrel{:=}-\frac{\bra{g}\textbf{d}\ket{e}E_0}{\hbar}$.
Thus, the Hamiltonian of a two level system driven by a near resonant laser in the rotating frame of the drive laser, using the rotating wave approximation is given by:
\begin{equation}
    \widetilde{H} = -\hbar\Delta_{la}\sigma^\dagger\sigma + \frac{\hbar \Omega_r}{2} \left( \sigma + \sigma^\dagger \right)
\end{equation}
Here, we have explicitly included the tilde to indicate the use of the rotating frame.
We define the state of the atom in the rotating frame as $\ket{\widetilde{\psi}} = c_g\ket{g} + \widetilde{c_e}\ket{e}$, where, $\widetilde{c_e} = c_e e^{i\omega_l t}$. We can then define the rotating frame density matrix $\widetilde{\rho}$ as $\ket{\widetilde{\psi}} \bra{\widetilde{\psi}}$.
The unitary evolution of the density matrix is given by:
\begin{equation}
    \partial_t \widetilde{\rho} = -\frac{i}{\hbar}[\widetilde{H}, \widetilde{\rho}]
\end{equation}

\begin{figure}[t]
    \begin{center}
    \includegraphics[width=0.9\textwidth, keepaspectratio]{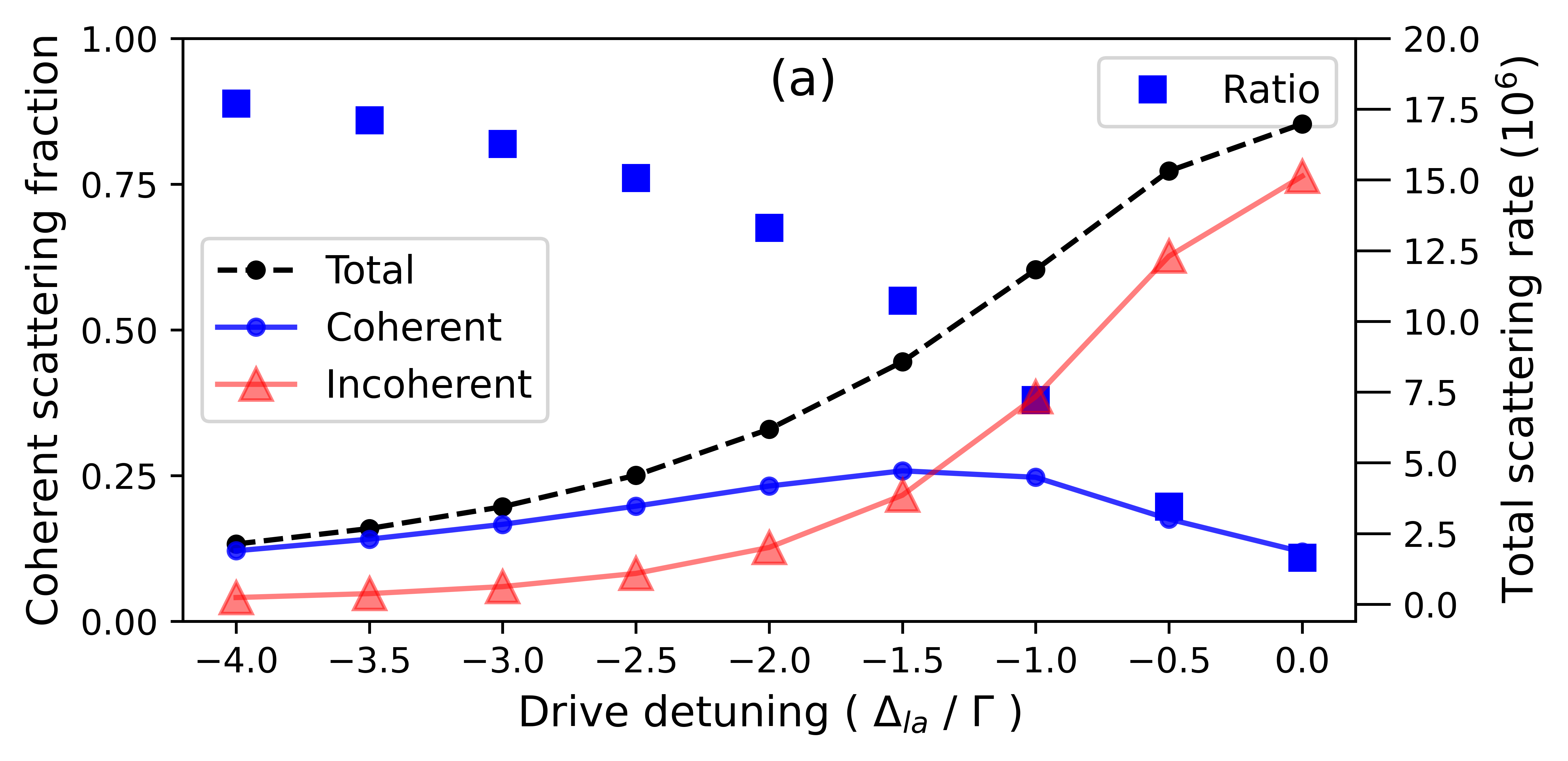}
        \label{Scattering}
        \caption{The plot shows the total, coherent and incoherent scattering rates for each atom versus drive laser detuning as black circles with dashed line, blue circles with solid line and red triangles respectively. The fraction of coherent to total scattering is presented as blue squares.}
	\end{center}
\end{figure}

Here, the square brackets indicate the commutator between the two operators. The components of the density matrix have the form $\widetilde{\rho}_{\alpha \beta} = \widetilde{c}_\alpha \widetilde{c}^*_\beta$, $\widetilde{c}_g = c_g$.
Let $\Gamma$ denote the spontaneous decay rate from the excited state. Since in our experiments, the atoms are confined in a magneto-optical trap, we ignore thermal and collisional broadening.
This system can be written in the form of coupled first order liner differential equations in the optical Bloch formalism to obtain:
{\setlength{\jot}{3ex}
\begin{align}
    \partial_t\widetilde{\rho}_{eg} &= -(\frac{\Gamma}{2}-i\Delta_{la})\widetilde{\rho}_{eg} + i\frac{\Omega_r}{2}\braket{\sigma_z}\\
    \partial_t\widetilde{\rho}_{ge} &= -(\frac{\Gamma}{2}+i\Delta_{la})\widetilde{\rho}_{ge} - i\frac{\Omega_r}{2}\braket{\sigma_z}\\
    \partial_t\braket{\sigma_z} &= i\Omega_r(\widetilde{\rho}_{eg} - \widetilde{\rho}_{ge}) - \Gamma(\braket{\sigma_z}+1)
\end{align}
}
Here, $\braket{\sigma_z} = \widetilde{\rho}_{ee} - \widetilde{\rho}_{gg}$ which gives the difference between the populations of the ground and excited states. The populations of the ground and excited states are the same in the lab frame and the rotating frame, i. e. $\rho_{ee} = \widetilde{\rho}_{ee}$ and $\rho_{gg} = \widetilde{\rho}_{gg}$. Thus, optical Bloch equations allow us to calculate the steady state population $\rho_{ee}(t\rightarrow\infty)$ and the coherence $\widetilde{\rho}_{eg}(t\rightarrow\infty)$ as \cite{Mollow_Steck}:
{\setlength{\jot}{3ex}
\begin{align}
    \rho_{ee}(t\rightarrow\infty) &= \frac{\frac{\Omega_r^2}{\Gamma^2}}{1+(\frac{2\Delta_{la}}{\Gamma})^2 + 2\frac{\Omega_r^2}{\Gamma^2}}; \\
    \widetilde{\rho}_{eg}(t\rightarrow\infty) &= -i\frac{\Omega_r}{\Gamma} \frac{1+\frac{2i\Delta_{la}}{\Gamma}}{1+(\frac{2\Delta_{la}}{\Gamma})^2 + 2\frac{\Omega_r^2}{\Gamma^2}}
\end{align}
}

Using the optical Wiener-Khinchin theorem and the definition of the dipole operator, the spectral function $S(\omega_s)$ of the scattered light in the rotating frame is given by:
\begin{equation}
    S(\omega_s) \mathrel{:=} \frac{1}{2\pi}\int_{-\infty}^{\infty}d\tau e^{i(\omega_s-\omega_l)\tau}\braket{\sigma^\dagger(t)\sigma(t+\tau)}
\end{equation}
Here, $\omega_s$ is the frequency of the scattered light. This allows us to calculate the total scattered intensity by integrating the spectral function over the entire frequency range as:
\begin{equation}
    \int_0^\infty S(\omega_s) d\omega_s = \braket{\sigma^\dagger\sigma}
\end{equation}
The expectation value of $\braket{\sigma^\dagger\sigma} = \rho_{ee}(t\rightarrow\infty)$. Finally, the total photon scattering rate is given by $R_{sc} = \Gamma \rho_{ee}(t\rightarrow\infty)$.
The spectrum of the scattered light can be divided into coherent and incoherent components. The coherence function $\braket{\sigma^\dagger(t)\sigma(t+\tau)}$ function decays to a constant non-zero value over large values of $\tau$.
This constant value which corresponds to square of the mean dipole moment of the atom, $|\widetilde{\rho}_{eg}(t\rightarrow\infty)|^2$, generates the coherent part of the spectrum of scattered light. The incoherent part of the scattered spectrum originates from the decaying part of the coherence function which is related to the fluctuations in the dipole moment.
In terms of the saturation parameter $s \mathrel{:=} \frac{2\Omega_r^2/\Gamma^2}{1+(2\Delta_{la}/\Gamma)^2}$ and the previously calculated steady state value of $\widetilde{\rho}_{eg}$, the total scattering rate along with the coherent and incoherent parts is given by:
{\setlength{\jot}{3ex}
\begin{align}
    R_{sc} &= \frac{\Gamma}{2}\frac{s}{1+s} \\ 
    R_{sc}^{coh} &= \frac{\Gamma}{2}\frac{s}{(1+s)^2} \\
    R_{sc}^{incoh} &= R_{sc} - R_{sc}^{coh} = \frac{\Gamma}{2}\frac{s^2}{(1+s)^2}
\end{align}
}
We show the variation of the total scattering rate along with the coherent and incoherent components as a function of drive laser detuning ($\Delta_{la}$) in Fig. S1.

\section{Near resonance Mollow spectrum}

\begin{figure}[t]
    \begin{center}
    \includegraphics[width=0.75\textwidth, keepaspectratio]{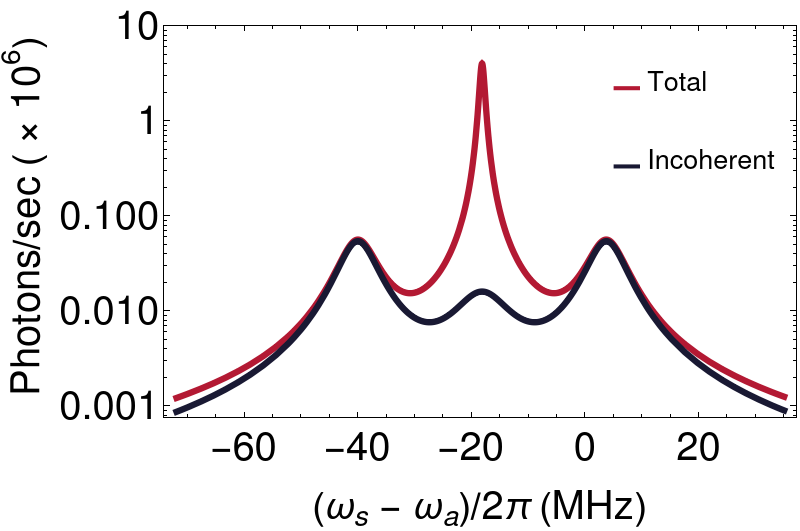}
        \label{Mollow}
        \caption{The plot shows the numerically calculated spectrum of incoherent part of scattered light in black and the total spectrum of scattered light in red. The intensity is denoted as the rate of scattered photons per second. The values of experimental parameters used to numerically solve the matrix equation are $\Delta_{la} \approx -3\Gamma$, $\Gamma/2\pi = 6.066$ MHz, $\Omega_r/2\pi \approx 12$ MHz. We add a Lorentzian with an FWHM (full width at half maxima) of 1 MHz to account for the linewidth of the drive laser, to obtain the complete spectrum of light scattered by a driven two level system.}
	\end{center}
\end{figure}

The coherent part of the spectrum generates a sharp peak at the drive laser frequency $\omega_l$. Therefore, to obtain the total spectrum, the incoherent part of the scattered light spectrum must be calculated. Since this part of the spectrum originates from the fluctuations in the dipole moment, we define fluctuation calculation operators $\delta\sigma^\dagger = \sigma^\dagger - \braket{\sigma^\dagger}_{t\rightarrow\infty}$ and $\delta\sigma = \sigma - \braket{\sigma}_{t\rightarrow\infty}$, where the subscript $t\rightarrow\infty$ indicates the steady state values of the operators. In the place of $\braket{\sigma^\dagger(t)\sigma(t+\tau)}$, we must now compute $\braket{\delta\sigma^\dagger(t)\delta\sigma(t+\tau)}_{t\rightarrow\infty}$. To calculate this modified coherence function, we must use the basic ideas of the quantum regression theorem \cite{Mollow1, Mollow_Steck, carmichael_lecture}.

Since we wish to compute the modified coherence function for the system in steady state, we use the quantum regression theorem in the long-time limit. This can be summarized as:
\begin{equation}
    \braket{A(t)B(t+\tau)} = \text{Tr}[B\Lambda(\tau)]
\end{equation}
Here, $A(t)$ and $B(t)$ are arbitrary operators in the Heisenberg picture.
The initial condition is given by $\Lambda(0)= \widetilde{\rho}(t\rightarrow\infty)A$ and its evolution can be calculated using the Liouvillian operator $\mathcal{L}$ as: $\partial_\tau\Lambda(\tau)=\mathcal{L}\Lambda(\tau)$.
Using the quantum regression theorem, to compute the modified coherence function, the initial condition is $\delta\Lambda(0)=\widetilde{\rho}(t\rightarrow\infty)\delta\sigma^\dagger$.
Instead of using the Liouvillian operator to calculate the evolution of $\delta\Lambda(\tau)$ we consider the optical Bloch equations. By subtracting the steady state part, the fluctuations in the density matrix elements can be cast into the following matrix equation \cite{Mollow_Steck, carmichael_lecture}:
\begin{equation}
    \partial_t\begin{bmatrix}
        \delta \widetilde{\rho}_{eg} \\ \delta \widetilde{\rho}_{eg} \\ \braket{\delta\sigma_z}
    \end{bmatrix} = \begin{bmatrix}
        -\frac{\Gamma}{2}+i\Delta_{la} & 0 & i\frac{\Omega_r}{2} \\
        0 & \frac{\Gamma}{2}-i\Delta_{la} & -i\frac{\Omega_r}{2} \\
        i\Omega_r & -i\Omega_r & -\Gamma
    \end{bmatrix}
    \begin{bmatrix}
        \delta \widetilde{\rho}_{eg} \\ \delta \widetilde{\rho}_{eg} \\ \braket{\delta\sigma_z}
    \end{bmatrix}
\end{equation}
The above equation is of the form: $\partial_\tau\delta\Lambda = \textbf{L}\delta\Lambda$, where $\delta\Lambda$ describes the fluctuations in the system and \textbf{L} acts similar to the Liouvillian operator in evolution part of the quantum regression theorem.
The initial condition $\delta\Lambda(0)$ can be calculated in component form as: $ \delta\Lambda_{\alpha\beta}(0) = \delta_{\beta g}\widetilde{\rho}_{\alpha e}(t\rightarrow\infty) - \widetilde{\rho}_{\alpha\beta}(t\rightarrow\infty)\widetilde{\rho}_{ge}(t\rightarrow\infty)$. In matrix form, this is:
\begin{equation}
    \delta\Lambda(0) = \begin{bmatrix}
        \rho_{ee} - |\widetilde{\rho}_{eg}|^2 \\
        -\widetilde{\rho}_{ge}^2 \\
        -\rho_{ee}\widetilde{\rho}_{ge} -\widetilde{\rho}_{ge} + \rho_{gg}\widetilde{\rho}_{ge}
    \end{bmatrix}_{t\rightarrow\infty}
\end{equation}
To obtain $\braket{\delta\sigma^\dagger(t)\delta\sigma(t+\tau)}_{t\rightarrow\infty}$ we need the component $\delta\Lambda_{eg}(\tau)$. We solve for this component numerically and compute the Fourier transform to obtain the spectrum of incoherent part of the scattered light.
Finally, to obtain the complete spectrum of the scattered light, we add to this a Lorentzian at the drive laser frequency. The total spectrum and the incoherent part are shown in Fig. S2.

\section{Collective Vacuum Rabi Splitting}

Let the resonant angular frequency of the empty cavity be denoted by $\omega_{c}$, $\omega_a$ be the angular frequency of atomic resonance and $\omega_p$ be the angular frequency of the probe laser. $a$ and $a^\dagger$ denote the photon annihilation and creation operators for the field inside the cavity.
The maximum single-atom-cavity interaction strength is calculated as $g_0 = -\mu \sqrt{\omega_c/(2\hbar\epsilon_0 V)}$, where, $V$ is the cavity mode volume and $\mu$ is the transition dipole matrix element for the relevant transition of the atom.
Due to the spread in the atomic cloud density, not all atoms interact with the cavity mode with the strength $g_0$. The interaction strength of the $j^{th}$ atom is given by $g_j = g_0 f(r_j)$. Here, $r_j$ is the position vector of the $j^{th}$ atom and the factor $f(r_j)$ is obtained from the geometrical mode function of the cavity field.
Let $\sigma_j^+$ and $\sigma_j^-$ be the atomic raising and lowering operators for the $j^{th}$ atom. Let $\Delta_{pa} = \omega_p - \omega_a$ and $\Delta_{pc}=\omega_p-\omega_c$ be the atom-probe detuning and probe-cavity detuning respectively.
Then, the time independent Hamiltonian of a classical probe laser field driving a system of N two-level atoms interacting with a cavity is given by in the rotating frame of the probe laser using the rotating wave approximation can be written as \cite{SAR_Rahul, rahul_thesis, Rempe_3photon}:
\begin{equation}
\hat{H} = \hbar \eta \left( \hat{a} + \hat{a}^\dagger \right) - \hbar \Delta_{pc} \hat{a}^\dagger \hat{a} + \sum_j^N(-\hbar \Delta_{pa} \hat{\sigma}^+ \hat{\sigma}^- + \hbar g_j \left( \hat{a}^\dagger \hat{\sigma_j}^- + \hat{a} \hat{\sigma_j}^+ \right))
\end{equation}
The equation of motion for a quantum mechanical observable $X$ in the Schr\"{o}dinger picture is given by:
\begin{equation}
\dot{X} = \frac{i}{\hbar} [H, X] + \sum_{i} \gamma_{i} \left( L_{i}^\dagger X L_{i} - \frac{1}{2} \left\{ L_{i}^\dagger L_{i}, X \right\} \right).
\end{equation}
Here, $L_i$ represents a collapse operator and $\gamma_j$ is the corresponding damping rate. Together this accounts for the spontaneous decay rate of the atomic excited state ($\Gamma$) and cavity photon loss rate from each mirror ($\kappa_t$).
Using the Lindblad-type evolution for elements of the density matrix, we can write the system in the form of coupled differential equations as:
\begin{align}
\frac{d\alpha(t)}{dt} &= -\left(\kappa_t - i\Delta_{pc}\right)\alpha(t) - i \sum_{j=1}^{N} g_j \rho_j^{eg}(t) - \eta \\
\frac{d\rho_j^{eg}(t)}{dt} &= -\left(\frac{\Gamma}{2} - i\Delta_{pa}\right)\rho_j^{eg}(t) + i g_j \alpha(t)\left(2\rho_j^{ee}(t) - 1\right)\\
\frac{d\rho_j^{ee}(t)}{dt} &= -\Gamma \rho_j^{ee}(t) + i \left[ g_j \alpha^*(t) \rho_j^{eg}(t) - g_j \alpha(t) (\rho_j^{eg}(t))^* \right]
\end{align}
Here, the classical probe laser field is is denoted by a coherent state $\ket{\alpha}$, the coherence of the $j^{th}$ atom is denoted by $\rho^{eg}_j$ and the excited state population of the $j^{th}$ is given by $\rho^{ee}_j$.
In the limit of weak probe laser field, assuming that most of the population stays in the ground state, one can obtain the steady state solution for the cavity field as:

\begin{equation}
  \alpha = 
\frac{-\eta (\left(\frac{\Gamma}{2} - i\Delta_{pa}\right)}{\left(\frac{\Gamma}{2} - i\Delta_{pa}\right)\left(\kappa_t - i\Delta_{pc}\right) + \sum_j^Ng_j^2}
\end{equation}

To simplify the expression for $\sum_j^Ng_j^2$, we take into account experimental conditions. In our setup, the atoms are confined in a magneto-optical trap (MOT). The MOT atoms are not strictly stationary. Thus, each atom experiences varying atom-cavity coupling. Hence, the average atom-cavity interaction for all atoms is identical. Let this value be $g$, such that $g<g_0$.
This allows us to write $\sum_j^Ng_j^2 = Ng^2$. This expression remains valid as long as the average number of atoms interacting with the cavity mode during our experiment remains constant. We now define $N_c$ as the effective number of atoms interacting with the cavity as $N_c g_0^2 = Ng^2$. That is, the effects produced by $N$ atoms interacting with an average coupling of $g$ will be identical to those of $N_c$ atoms, all interacting with the cavity with maximum coupling $g_0$. Thus, we get the final form of the cavity field amplitude as:
\begin{equation}
  \alpha = 
\frac{-\eta (\left(\frac{\Gamma}{2} - i\Delta_{pa}\right)}{\left(\frac{\Gamma}{2} - i\Delta_{pa}\right)\left(\kappa_t - i\Delta_{pc}\right) + N_cg_0^2}
\end{equation}
We use this equation to calculate the vacuum Rabi splitting produced by MOT atoms confined inside the cavity mode at relevant cavity detunings. The expected VRS signal of an arbitrarily weak probe scanning across the D2 $F=3 \leftrightarrow F'=4$ transition of $^{85}$Rb for $N_c =23000$ when the cavity is tuned to the D2 $F=3 \leftrightarrow F'=4$ transition ($\omega_c = \omega_a$) is shown in Fig. S3.

\begin{figure}[t]
    \begin{center}
    \includegraphics[width=0.75\textwidth, keepaspectratio]{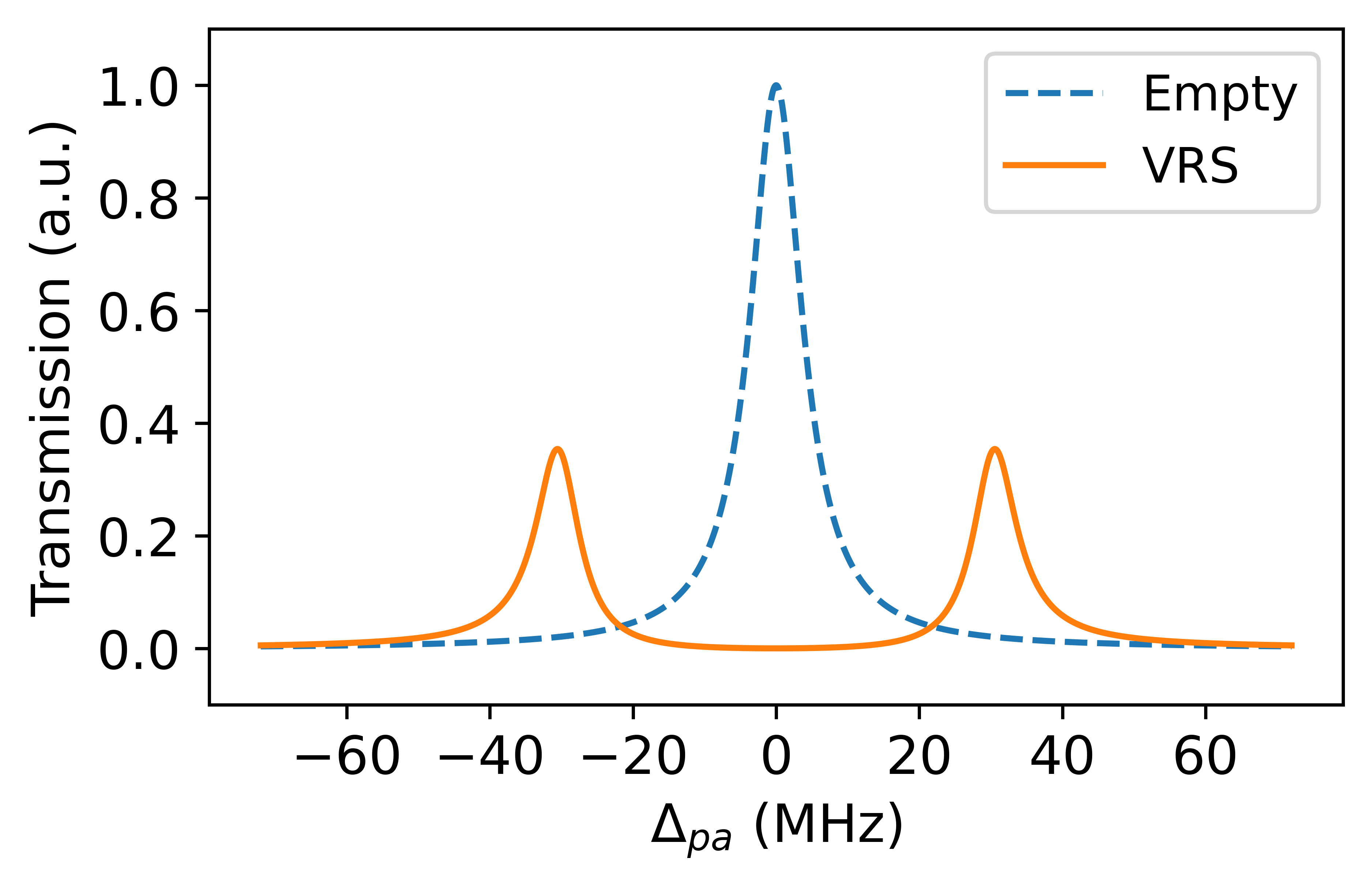}
        \label{VRS}
        \caption{The plot shows the transmission of a weak probe through an empty cavity as the dashed blue curve. The solid curve shows the cavity transmission of the weak probe with $N_c =23000$ and $\omega_c = \omega_a$. The transmission intensities are normalized with respect to the peak of empty cavity transmission.}
	\end{center}
\end{figure}

\bibliography{supplemental}